\documentclass[a4paper,twocolumn,showpacs,preprintnumbers,amsmath,amssymb,nofootinbib,floatfix]{revtex4}
\usepackage{hyperref}
\input psfig.sty
\usepackage{picinpar}
\usepackage{graphicx}

\begin{document}

\title{Smoothing expansion rate data to reconstruct cosmological matter perturbations}

\author{J. E. Gonzalez\footnote{E-mail: javierernesto@on.br}}
\author{J. S. Alcaniz\footnote{E-mail: alcaniz@on.br}}
\author{J. C. Carvalho\footnote{E-mail: jcarvalho@on.br}}

\affiliation{Departamento de Astronomia, Observat\'orio Nacional, 20921-400, Rio de Janeiro - RJ, Brasil}

\date{\today}

\begin{abstract}

The existing degeneracy between different dark energy and modified gravity cosmologies at the background level may be broken by  analysing  quantities at the perturbative level. In this work, we apply a non-parametric smoothing  (NPS) method to reconstruct the expansion history of the Universe ($H(z)$) from model-independent cosmic chronometers and high-$z$ quasar data. Assuming a homogeneous and isotropic flat universe and general relativity (GR) as the gravity theory, we calculate the non-relativistic matter perturbations in the linear regime using the $H(z)$ reconstruction and realistic  values of $\Omega_{m0}$ and $\sigma_8$ from Planck and WMAP-9 collaborations. We find  a good agreement between the  measurements of the growth rate and $f\sigma_8(z)$ from current large-scale structure observations and the estimates obtained from the reconstruction of the cosmic expansion history. Considering a recently proposed null test for GR using matter perturbations, we also apply the NPS method to reconstruct $f\sigma_8(z)$. For this case, we find a $\sim 2\sigma$ tension (good agreement) with the standard relativistic cosmology when the Planck (WMAP-9) priors are used.

\end{abstract}

\maketitle

\section{Introduction}
The discovery of cosmic acceleration poses a very fundamental question to theoretical cosmology: does acceleration reflect the existence of new fields in high energy physics or the need for modifications of the standard gravity theory on cosmological scales? This question becomes even more complicated as it is  possible to construct modified gravity (MG) scenarios that produce the same cosmic expansion of general relativistic dark energy (DE) models~\cite{Wang2008}. Such degeneracy lies in the background level, where the description of the Universe is made from a perfectly symmetric and simplified model. Therefore, the use of cosmological observables, such as the Hubble parameter or the luminosity or angular diameter distances, does not seem to be enough to determine if the cosmic acceleration is a geometrical or a dynamical effect. When a more realistic scenario is considered, in which the universe has geometrical and energy fluctuations, it is possible to describe the growth and evolution of overdensity and underdensity regions. Since different classes of models produce, in general, characteristic predictions of the growth of the cosmic structures, perturbative quantities, such as the  growth rate ($f$) and index ($\gamma$), are believed to be important tools to distinguish MG from DE models (see, e.g., \cite{Wang2008,GannoujiPolarski,PolarskiGannouji}).

From the theoretical side, some null tests have been proposed to probe the validity of the standard cosmology \cite{Sahni_Omtest,Clarkson_Omtest,Clarkson_Oktest}. In these tests, the relation between different observables must be set at specific values, otherwise, there would be a violation of one or more assumptions used to derive the test. In Refs. \cite{Nesseris2014, Nesseristest} it was presented a new null test involving both measurements of cosmic expansion and matter perturbations. For the arguments mentioned earlier, it is expected that such null tests involving perturbative observables will be more efficient than others using only background quantities.

On the other hand, to obtain information about the functional behaviour of dynamical and kinematical variables, parametric and non-parametric methods have been used. In both cases, the final reconstructed quantity is constrained by observational data. In parametric approaches 
(see, e.g., \cite{BAZR,BA,CPL-Polarski,CPL-Linder,Jassal}), a prior functional form is used to describe the observations, 
whereas in the non-parametric approaches (see, e.g., \cite{Sendra,Vitenti,Montiel,Maturi,Mignone,Herrera,Rasmussen,Seikel,Seikel2012,Shafieloo2006,Shafieloo2012a,Li,Gonzalez}), 
it is commonly assumed a correlation between each data point.

In this work, we apply a non-parametric smoothing (NPS) method \cite{Shafieloo2006,Shafieloo2007,Shafieloo2012a,Li,Gonzalez} to reconstruct the evolution of matter perturbations from background data, such as the measurements of the cosmic expansion rate $H(z)$. We apply the method to a sample of  $H(z)$ data from cosmic chronometers~\cite{Alcaniz:2001uy, Jimenez03,Simon05,Stern10,Moresco12,Moresco:2015cya}, lying in the redshift  $0.070 \leq z \leq 1.2$, and high-$z$ quasar data at $z\approx 2.3$ \cite{Busca}.  The cosmic chronometer data have been obtained from the differential age method for passively evolving galaxies of Ref.~\cite{Jimenez-Loeb},  which is aimed to be cosmological and stellar population synthesis model-independent \cite{Licia}. On the other hand, current measurements of the expansion rate from quasar data have been obtained using the three-dimensional correlation function of the transmitted flux fraction in the Ly$\alpha$-forest of high-$z$ quasars, as reported in Ref.~\cite{Busca}. In particular, the application of this latter technique to a sample of 48,640 quasars provided a measurement of $H(z)$ within $\sim 3\%$ accuracy at $z = 2.3$, which imposes tight bounds on cosmological parameters when combined with current $H_0$ measurements and other cosmological data sets (see, e.g., \cite{ratra} for a recent analysis). We follow the approach presented in Ref.~\cite{Sahni2009} (see also \cite{Gonzalez}) to reconstruct perturbative quantities from background observables and compare them with current measurements of growth rate and index.  Finally, we also use the NPS method to reconstruct the $f\sigma_8(z)$ observable and evaluate the  null test proposed in Ref.~\cite{Nesseris2014,Nesseristest}. 

This paper is organised as follows: in Sec \ref{MPE} we summarise the treatment of linear matter perturbations of Ref. \cite{Sahni2009}
and the perturbative null test proposed in Ref.~\cite{Nesseris2014,Nesseristest}.  We also introduce the basic equations of the matter perturbation theory and a brief explanation on how to construct the null test. In Sec. \ref{DaHPR} we present the observational data and the non-parametric method used to reconstruct the evolution of $H(z)$ and the perturbative quantities. We present the results of our reconstructed cosmic expansion, the matter perturbation analysis and the calculation of the null test in Sec. \ref{R}.  We end this paper with the main conclusions in Sec. \ref{C}.

\section{Matter Perturbation Equations}

\subsection{Matter Perturbation Description}
\label{MPE}

In the longitudinal gauge, the scalar perturbations of a flat FLRW metric are characterised by the line element
\begin{equation}
 ds^2=(1+2\Phi)dt^2-(1-2 \Psi)a^2(t)d \vec{x} ^2\;,
\end{equation}
where $\Phi$ and $\Psi$ correspond to the gauge invariant potential and curvature perturbation, respectively. When we 
consider scales inside the Hubble sphere, the matter density contrast,
\begin{equation}
 \delta(\vec{x},t)\equiv \frac{\rho(\vec{x},t)-\rho(t)}{\rho(t)},
\end{equation}
and the scalar modes satisfy the Poisson equation \footnote{In GR the scalar modes $\Phi$ and $\Psi$ are equal if there is not anisotropic stress in the Universe components. }
\begin{equation}
 \nabla^2 \Phi =4\pi G a^2 \rho_m \delta\;.
\end{equation}
We consider  a universe composed of matter and an unclustered DE fluid separately conserved. Under these assumptions, the evolution of the matter density contrast is given by~\cite{Starobinsky1998} 
\begin{equation}
\label{2ode}
\ddot{\delta}+2H\dot{\delta}-4 \pi G \rho_m \delta = 0\;,
\end{equation}
where the dot denotes derivative with respect to cosmic time.

 We can solve Eq.(\ref{2ode}) in terms of the follow set of integral equations \cite{SahniStarobinsky,Sahni2009} 
\begin{subequations}
\begin{eqnarray}
\label{22odes}
\delta(D)&=&1+\delta_0 '\int_0 ^D[1+z(D_1)]dD_1  \\ &&
 +\frac{3}{2}\Omega_{m0}\int_0^D[1+z(D_1)]\left(\int_0^{D_1}\delta(D_2)dD_2  \right)dD_1 \nonumber \;,
\end{eqnarray}
\begin{eqnarray}
 \label{22odes2}
\delta'(D)&=&\delta_0 '[1+z(D)] \\ && +\frac{3}{2}\Omega_{m0}[1+z(D)] 
\int_0^{D}\delta(D_1)dD_1 \nonumber \;.
\end{eqnarray} 
\end{subequations}
where $D$ represents the adimensional physical distance, defined by
\begin{equation}
D=H_0 \int_t^{t_0}\frac{dt}{a(t)}=H_0\int_0^z \frac{dz_1}{H(z_1)}\;.
\end{equation}
The prime denotes the derivative  with respect to $D$ and $\Omega_{m0}$ is the matter density parameter at $z=0$.
Note that the integral solution of the matter density contrast is not coupled with its first derivative. 
Conversely, to solve $\delta'$ we need to know the behaviour of the density contrast. 

The only prior information required to solve Eqs (\ref{2ode}) and (\ref{22odes}) is the value of the density parameter $\Omega_{m0}$. In this work, we use realistic values of the density parameter provided by current CMB experiments. To obtain a unique solution for the second order differential equation (\ref{2ode}), we need to fix the two integration constants. In the solution (\ref{22odes}) these constants are the values
of the matter density contrast and its first derivative at the present time. $\delta(z=0)$ is fixed by the  normalisation of  (\ref{22odes}) at $z=0$ whereas the second constant, $\delta'(z=0)=\delta'_0$, can be fixed 
analysing the behaviour of the solution at very high redshift, where it is expected that $\delta \propto a$. For this 
purpose, it is easier to analyse  another relevant perturbative quantity,  the so-called growth factor, defined by \cite{Sahni2009}
\begin{equation}
g(z)\equiv(1+z)\delta(z)\;.
\end{equation}
With this definition, we can fix the $\delta'_0$ when the growth factor reaches a constant behaviour at high $z$ (see \cite{Gonzalez} for more details).  The integral solution  (\ref{22odes}) of $\delta$  involves only information about the matter density parameter at the current epoch 
and the normalized expansion rate ($H(z)/H_0$), which is contained in the dimensionless physical distance\footnote{The matter density contrast solution is independent of the Hubble constant value.}. Prior knowledge of these two quantities and of $H(z)/H_0$ determine univocally the matter density contrast \cite{Wang2008, Nesseris2015}.

We can define the growth rate of the linear density contrast as
\begin{equation}
 f(z)\equiv\frac{d\ln \delta}{d\ln a}=-\frac{(1+z)H_0}{H(z)}\frac{\delta'}{\delta}\;.
 \label{fdef}
\end{equation}
and the rate of change of the amplitude of clustering,
\begin{equation}
 f\sigma_8(z) \equiv f(z)\sigma(z)=\frac{d \sigma_8(a(z))}{d\ln a},
 \label{fs8def}
\end{equation}
where 
\begin{equation}
 \sigma_8(z)=\sigma_{8}\delta(z)/\delta(0)
\end{equation}
is the rms
amplitude of mass fluctuations  in terms of $\delta(z)$ \cite{Nesseris2014}
and $\sigma_{8}$ is the present linear mass dispersion on a sphere of radius $8h^{-1}$Mpc. These two quantities can be inferred from the observations of the large-scale structure by analyzing the matter power spectrum  or from the gravitational lensing data \cite{Huterer, Reyes2010}. 

The measurements of $f(z)$ and $f\sigma_8(z)$  from current large-scale structure observations and the estimates obtained from Eqs. (\ref{2ode}), (\ref{fdef}) and (\ref{fs8def}) along with the reconstruction of the Hubble parameter are independent. A tension between these independent estimates would be a clear evidence of non-standard cosmology or a violation of the assumptions used to derive Eq.(\ref{2ode}). For instance:

\begin{itemize}
 \item The Universe is not flat, homogeneous and isotropic in large scales. For instance, in a non-flat universe the Poisson equation has to be modified, as shown in Ref. \cite{Marteens}.
 
 \item The DE and matter components are not separately conserved and  the matter density does not decrease proportional to  $a^{-3}$. Examples of this kind are models with decaying of dark energy into dark matter or vice versa \cite{Rsousa}. 
 
 \item The correct theory of gravity is not the GR or the DE fluid is clustering. In  both cases, $G$ is no longer  a  constant and can be written as  an effective gravitational function  which  depends on the scale and time $G_{eff}=G_{eff}(k,t)$ \cite{GannoujiPolarski, carroll, Tsujikawa,Mehrabi2}.
\end{itemize}
We refer the reader to Ref.~\cite{Sahni2009} for a more detailed discussion of this topic.

Finally, we  define the growth index, which we write as a function of $z$ as~\cite{Peebles,Lahav,Wang1998}
\begin{equation}
\gamma = \frac{d\ln{f(z)}}{d\ln{\Omega_{m}(z)}}\;,
 \label{fapprox}
\end{equation}
where
\begin{equation}
 \Omega_m(z) \equiv \frac{\Omega_{m0}(1+z)^3H_0^2}{H^2(z)}\;.
 \label{om}
\end{equation}
The main importance of the growth index is that it constitutes a powerful tool to characterize gravity theories. For instance,
the $\Lambda$CDM model is characterised by $\gamma=6/11$ and, in general,  slow varying DE models are predict  $\gamma \simeq \frac{3(w-1)}{6w-5}$ \cite{Wang1998,LinderCahn}. In other scenarios, like DGP model, $\gamma=11/16$ \cite{LinderCahn} whereas for the $f(R)$ gravity model discussed in Ref. \cite{Starobinsky2007} the growth index is constrained to the interval $0.40<\gamma<0.43$ \cite{GannoujiPolarski}.

\subsection{Null Test for GR \label{nt}}

It is possible to construct a null test for GR from matter density perturbations using the equations that govern their evolution. 
For this reason, the hypotheses used so that Eq.(\ref{2ode})  correctly describes the evolution of $\delta(z)$ must be satisfied. The null test is constructed as follows~\cite{Nesseris2014,Nesseristest}. First, let us assume that Eq.(\ref{2ode}) results from the Euler-Lagrange equations with $\delta$ as generalised coordinate, i.e., 
\begin{equation}
\frac {d}{da} \left(\frac{\partial{\cal{L}}(a,\delta,\dot{\delta})}{\partial \dot{\delta}}\right) -\frac{\partial{\cal{L}}(a,\delta,\dot{\delta})}{\partial {\delta}}=0.
\label{E-L}
\end{equation}
In this case, we can associate the follows Lagrangian ${\cal{L}}$ and Hamiltonian ${\cal{H}}$ to the system:
\begin{subequations}
\begin{eqnarray}
{\cal{L}}=\frac{1}{2}a^3\delta'(a)^2H(a)/H_0+\frac{3\Omega_{m0}}{4a^2H(a)/H_0}\delta(a)^2,
 \label{lagrangiano}
\end{eqnarray}
\begin{eqnarray}
{\cal{H}}=\frac{1}{2}a^3\delta'(a)^2H(a)/H_0-\frac{3\Omega_{m0}}{4a^2H(a)/H_0}\delta(a)^2.
 \label{hamiltoniano}
\end{eqnarray} 
\end{subequations}
In this system, the Hamiltonian depends explicitly on the 'time' parameter $a$ which implies that ${\cal{H}}$ is not a constant of 'motion'. 
Finding a symmetry direction of the system, we construct the first integral of motion,
\begin{equation}
 \Sigma=a^3H(a)/H_0\delta'(a)e^{-\int^a_{a_0}\frac{3\Omega_{m0}\delta(x)}{2x^5H(x)^2/H_0^2\delta'(x)}dx}.
 \label{test}
\end{equation}
This normalized quantity constitutes a null test of GR involving a perturbative variable. It is still possible to rewrite Eq.(\ref{test}) in terms of the observables $H(z)$ and $f\sigma_8(z)$, i.e. (considering $a_0=1$ and $G_{eff}=G$)
\begin{eqnarray}
\label{nulltest}
  {\cal{O}}(a)=\frac{a^2H(a)f\sigma_8(a)}{H_0f\sigma_8(1)} \times 
  \end{eqnarray}
  \begin{eqnarray*}
   \exp \left({-\frac{3}{2}\Omega_{m0}  \int^a_{1}\frac{\sigma_8(a=1)+\int^x_1\frac{f\sigma_8(y)}{y}dy}{x^4H(x)^2/H_0^2f\sigma_8(x)}}\right). &
 \end{eqnarray*}
 In addition to the assumptions used to obtain Eq.(\ref{2ode}) and its solution (\ref{22odes}), this test requires that there will not be tension between the cosmic expansion and matter density growth data. If these hypotheses are satisfied, the value of ${\cal{O}}(z)$ must be unity. 

 \begin{table}
	\begin{center}
 		\begin{tabular}{lcc}
 		\hline\hline
 $z$ & $H_{obs}(z)$ [km s$^{-1}$ Mpc$^{-1}$] & Ref. \\
 \hline
 0.100 & 69 $\pm$ 12 & \cite{Simon05} \\
 0.170 & 83 $\pm$ 8 & \cite{Simon05} \\
  0.179 & 75 $\pm$ 4 & \cite{Moresco12}\\
  0.199 & 75 $\pm$ 5 & \cite{Moresco12} \\
  0.270 & 77 $\pm$ 14 & \cite{Simon05} \\
  0.352 & 83 $\pm$ 14 & \cite{Moresco12} \\
  0.400 & 95 $\pm$ 17 & \cite{Simon05} \\
  0.480 & 97 $\pm$ 62 & \cite{Stern10} \\
  0.593 & 104 $\pm$ 13 & \cite{Moresco12} \\
  0.680 & 92 $\pm$ 8 & \cite{Moresco12} \\
  0.781 & 105 $\pm$ 12 & \cite{Moresco12}\\
  0.875 & 125 $\pm$ 17 & \cite{Moresco12} \\
  0.880 & 90 $\pm$ 40 & \cite{Stern10} \\
  0.900 & 117 $\pm$ 23 & \cite{Simon05} \\
  1.037 & 154 $\pm$ 20 & \cite{Moresco12} \\
 2.34 & 222 $\pm$ 7 & \cite{Debulac} \\
 2.36 & 226 $\pm$ 8 & \cite{Font-Ribera}\\
 \hline\hline
 		\end{tabular}
 	\end{center}
 	\caption{$H(z)$ data from 15 cosmic chronometer systems and  two high-$z$ quasar data.}
 	\label{Hdados}
 \end{table}

\section{Data and Hubble Parameter Reconstruction}
\label{DaHPR}

\subsection{Data}
\label{D}
Measurements of the expansion rate  are important observational probes of the late-time cosmic acceleration. The cosmic chronometer approach is developed using the differential ages of massive and passively evolving old elliptical galaxies~\cite{Jimenez-Loeb}. In contrast to other observables that depend on integrated quantities along the line of sight, cosmic chronometers estimates are independent of the spatial geometry of the Universe. 
The cosmic expansion corresponds to the estimate of the age change of the Universe for a given variation of the redshift. This information can be obtained by the analysis of the galaxy ages with their respective $z$. Considering passively evolving galaxies at approximately the same redshift, it is possible to obtain an estimate of the cosmic expansion using the expression $H(z)=-1/(1+z)\Delta z/\Delta t$. Here, $\Delta z$ is the difference between the redshift of two galaxies and $\Delta t$ the difference between their ages.

One of the main feature of cosmic chronometer approach lies in the fact that their estimates are cosmological model-independent, although  there can be dependence on stellar population synthesis models at high redshift. As emphasized in Ref. \cite{Licia}, this latter dependence does not appear until $z\simeq1.2$. For this reason, we use only 15 measurements of the $H(z)$ from cosmic chronometers, up to $z = 1.04$. As discussed in Ref. \cite{Licia} we  increase slightly (20\%) the error bar of the highest-$z$ data point to account for the uncertainties of the stellar population synthesis models. We also add to our sample two measurements of $H(z)$ from quasar data at very high-$z$, i.e., $z = 2.34$ \cite{Debulac} and $z = 2.36$ \cite{Font-Ribera}, which were obtained by determining the BAO scale from the correlation function of the Ly$\alpha$-forest systems. The complete data set used in our analysis is presented in Table \ref{Hdados}.

\begin{figure*}[t]
\includegraphics[width = 8.5cm, height = 6.3cm]{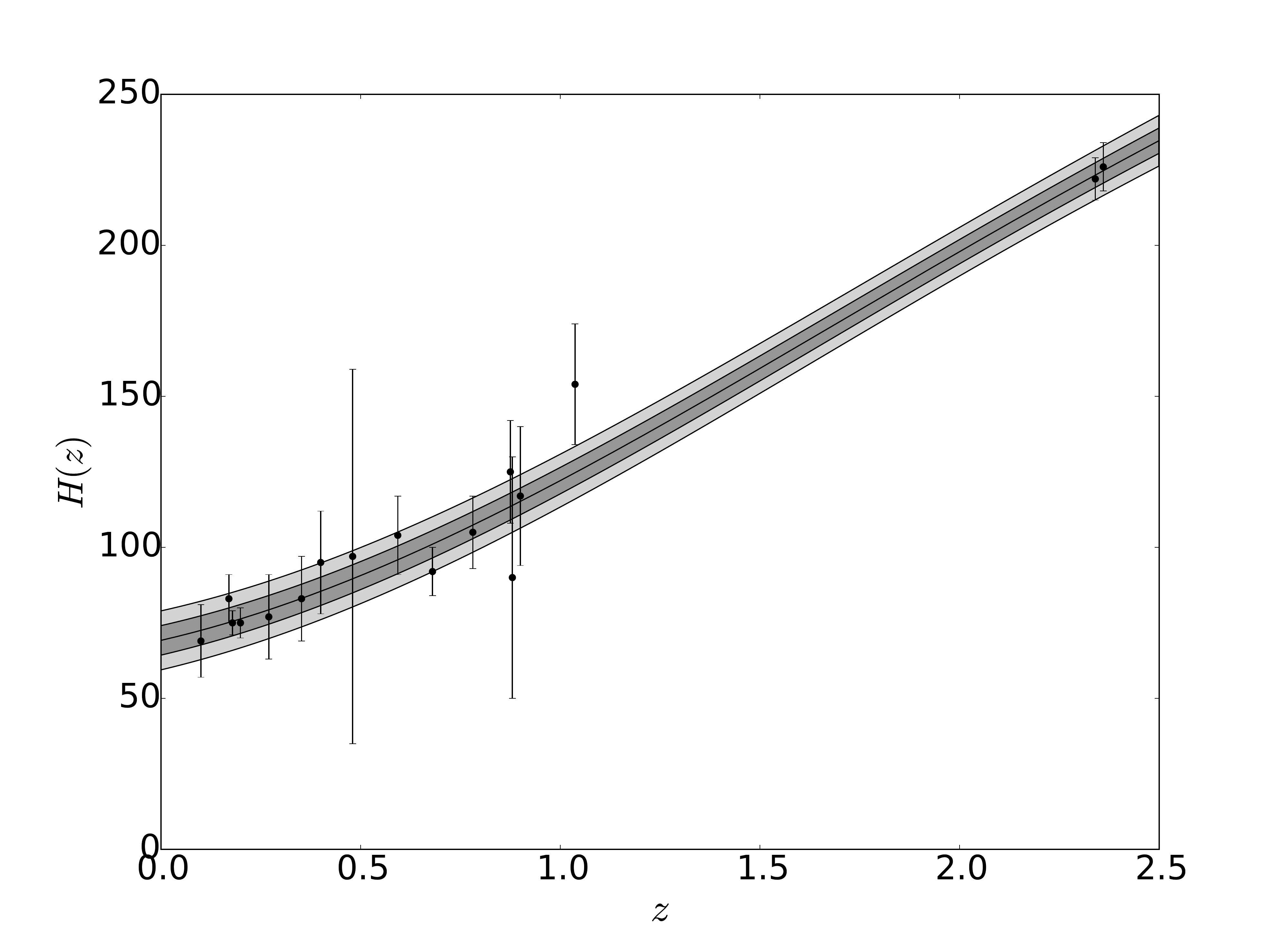}
\hspace{0.1cm}
\includegraphics[width = 8.5cm, height = 6.3cm]{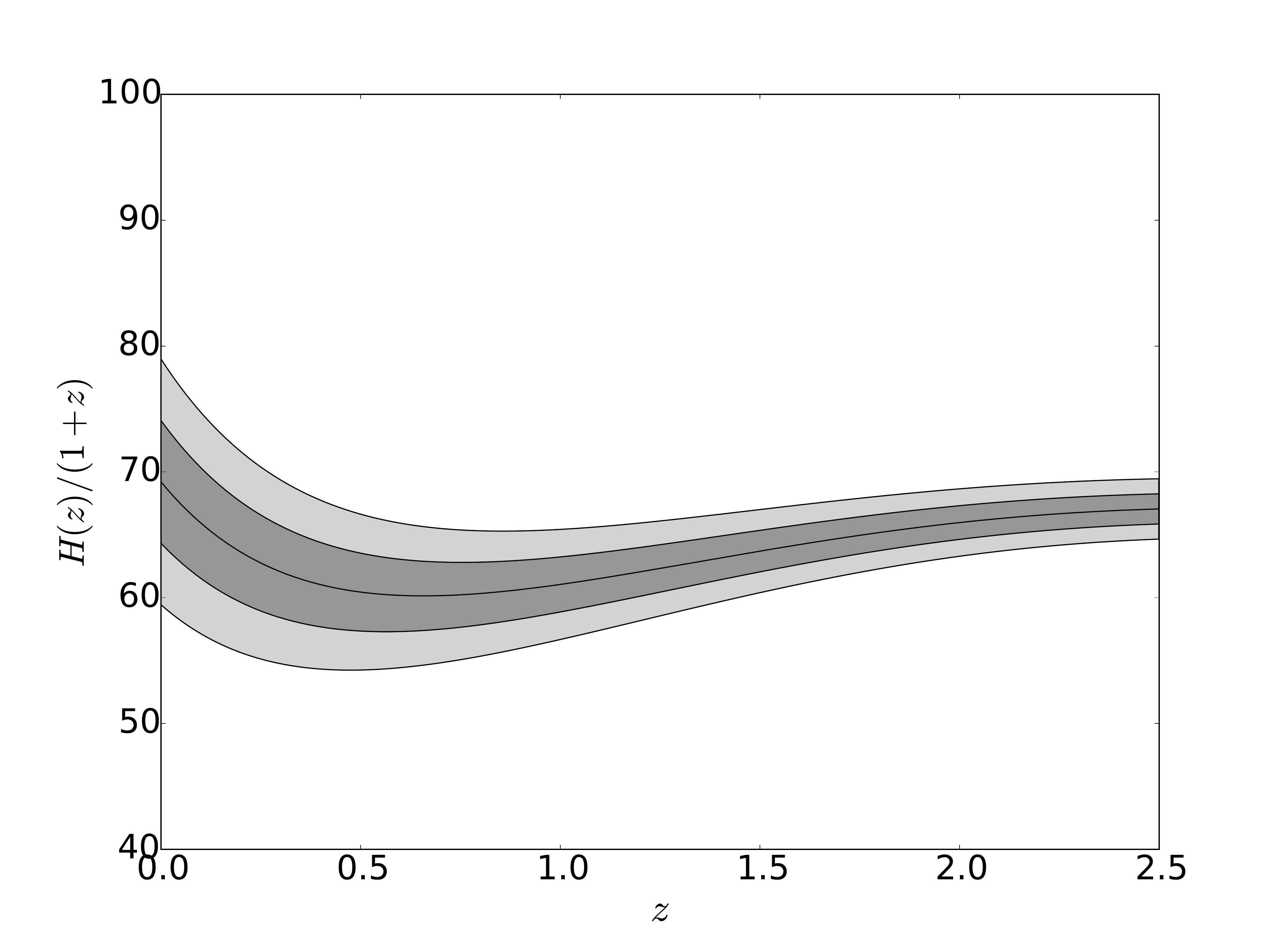}
\caption{(a) Reconstruction of the cosmic expansion (in km s$^{-1}$ Mpc$^{-1}$) via the NPS method using the cosmic chronometer and high-$z$ quasar data. 
The black solid line corresponds to the NPS reconstruction whereas the shaded regions to the 1$\sigma$ and 2$\sigma$ confidence intervals. The data points represent the observational data displayed in Table \ref{Hdados}. (b) The quantity $\dot{a}=H(z)/(1 + z)$ as a function of $z$.}
\label{Hzfig}
\end{figure*}

\begin{figure*}[t]
\includegraphics[width = 8.5cm, height = 6.4cm]{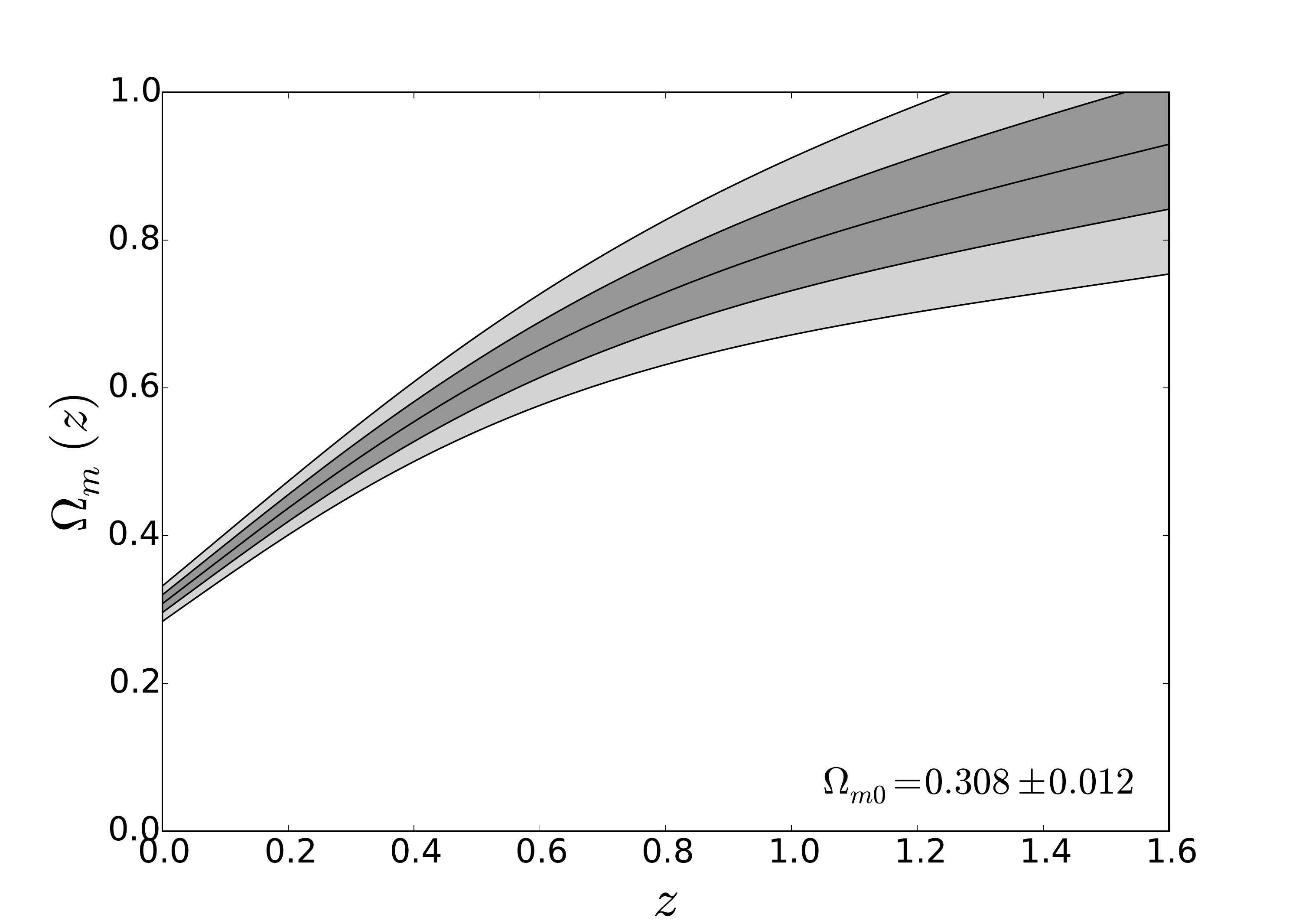}
\hspace{0.1cm}
\includegraphics[width = 8.5cm, height = 6.4cm]{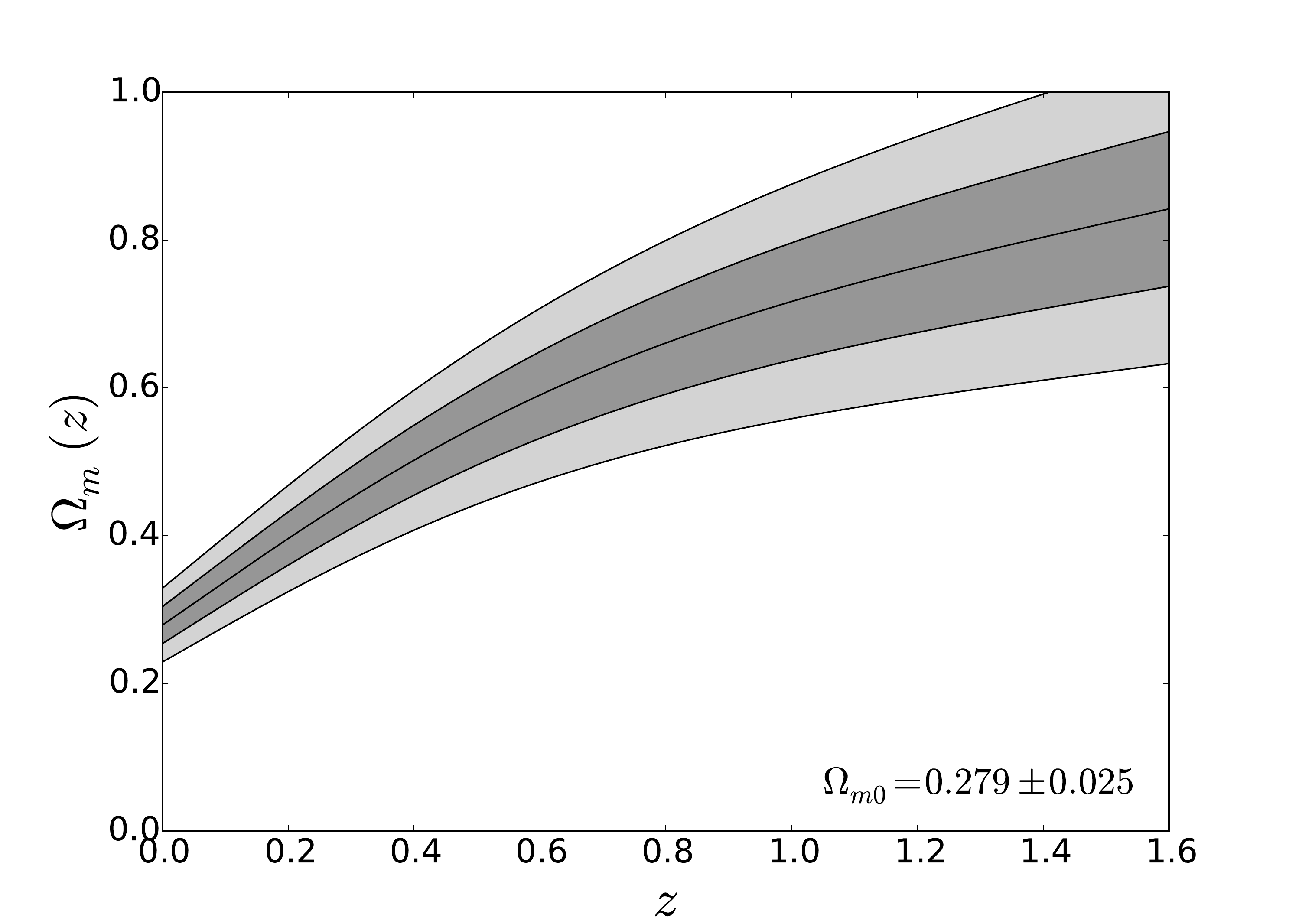}
\caption{The evolution of the matter density parameter calculated using the reconstruction of $H(z)$ shown in Fig. \ref{Hzfig}(a) and current estimates of $\Omega_{m0}$ from the Planck collaboration (a) and from the WMAP-9 collaboration (b). The shaded regions correspond to the 1$\sigma$ and 2$\sigma$ confidence intervals.}
\label{Omfig}
\end{figure*}

\subsection{Non-parametric Smoothing}
\label{NPS}
 In order to perform a model-independent reconstruction of the expansion rate 
 of the Universe from the cosmic chronometer  and high-z quasar data, we applied  
 the non-parametric method proposed in Ref. \cite{Shafieloo2006,Shafieloo2007,Shafieloo2012a,Gonzalez}. This method has been very useful in the reconstruction 
 of the luminosity and the physical distances and the Hubble parameter.
In its general form,  and taking into account the data errors, the smoothing quantity is obtained as
\begin{eqnarray}
H^s(z,\Delta) & = & H^g(z) + N(z) \sum_i \frac{\left[H(z_i)- H^g(z_i)
\right]} {\sigma^2_{H(z_i)}} \nonumber \\ & &  \times {\cal{K}}(z, z_i),
\end{eqnarray}
where $H^s(z,\Delta)$ is the smoothed quantity,  $H^g(z_i)$ is the 
initial guess model, $H(z_i)$ is the observational data, $\sigma_H(z_i)$ is the 
data error, $\Delta $ is the smoothing scale and $N(z)$ is the 
normalization factor given by:
\begin{equation}
 N(z)^{-1}= \sum_i {{\cal{K}}(z, z_i)\over {\sigma^2_{H(z_i)}}}.
\end{equation}
Following the procedure presented in Ref. \cite{Li}, in this work we adopt a Gaussian kernel (${\cal{K}}(z, z_i)=\exp(-(z-z_i)^2/2\Delta^2)$) to perform the reconstruction. 
Due to the iterative application of the smoothing function, the dependence on the initial guess model becomes insignificant and the reconstruction is 
effectively model-independent (see, e.g. \cite{Shafieloo2006,Shafieloo2007,Shafieloo2012a,Li,Gonzalez}).

In order to calculate the $1\sigma$ confidence level we use the approach developed in Ref. \cite{Bowman_smoothing} (see also \cite{Gonzalez}). In this case the error is given by:
\begin{equation}
 \sigma_{H^s(z)}= \left( \sum_i v^2_i \hat{\sigma}^2 \right)^{1/2},
 \label{1sigma}
\end{equation}
 $\sigma_{H^s(z)}$ being the $1\sigma$ confidence level of $H(z)$, $v_i$  the smoothing factor ($v_i=N(z) {\cal{K}}(z,z_i)/\sigma^2_{H(z_i)}$)
and $\hat{\sigma}^2$ is the estimate of the error variance given by \cite{Li,Bowman_smoothing}:

\begin{equation}
 \hat{\sigma}^2=\frac{\sum_i \left(H(z_i)-H^s(z_i) \right)^2}{\sum_j (1-v_j(z_j))}.
\end{equation}
As done in Ref. \cite{Li} we increase the value of the $1\sigma$ confidence level ($\sigma_{H^s(z)}$) by 30$\%$.

The final reconstruction depends on the smoothing scale. For instance, for very small values of $\Delta$, the reconstructed curve tends to follow more closely the data points with several  bumps. On the other hand, if  $\Delta$ is too large, the curve is too smooth, departing significantly from the data. 
For this reason it is convenient to estimate an optimal value of $\Delta$. We select the $\Delta$ value that minimizes the cross-validation function  given by

\begin{equation}
 CV(\Delta)=\frac{1}{n}\sum_i (H(z_i)-H^s_{-i}(z_i|\Delta))^2,
\label{CV}
\end{equation}
where $H^s_{-i}(z_i|\Delta)$ denotes the reconstructed expansion at $z=z_i$ without taking into account the data point
$(z_i,H(z_i))$ for a given $\Delta$. For the cosmic chronometer and high-$z$ quasar data (see Sec. \ref{D}) the $\Delta$ value that asymptotically minimizes Eq.
(\ref{CV}) is $\sim 1.4$.

\begin{figure*}[t]
\includegraphics[width = 8.5cm, height = 6.5cm]{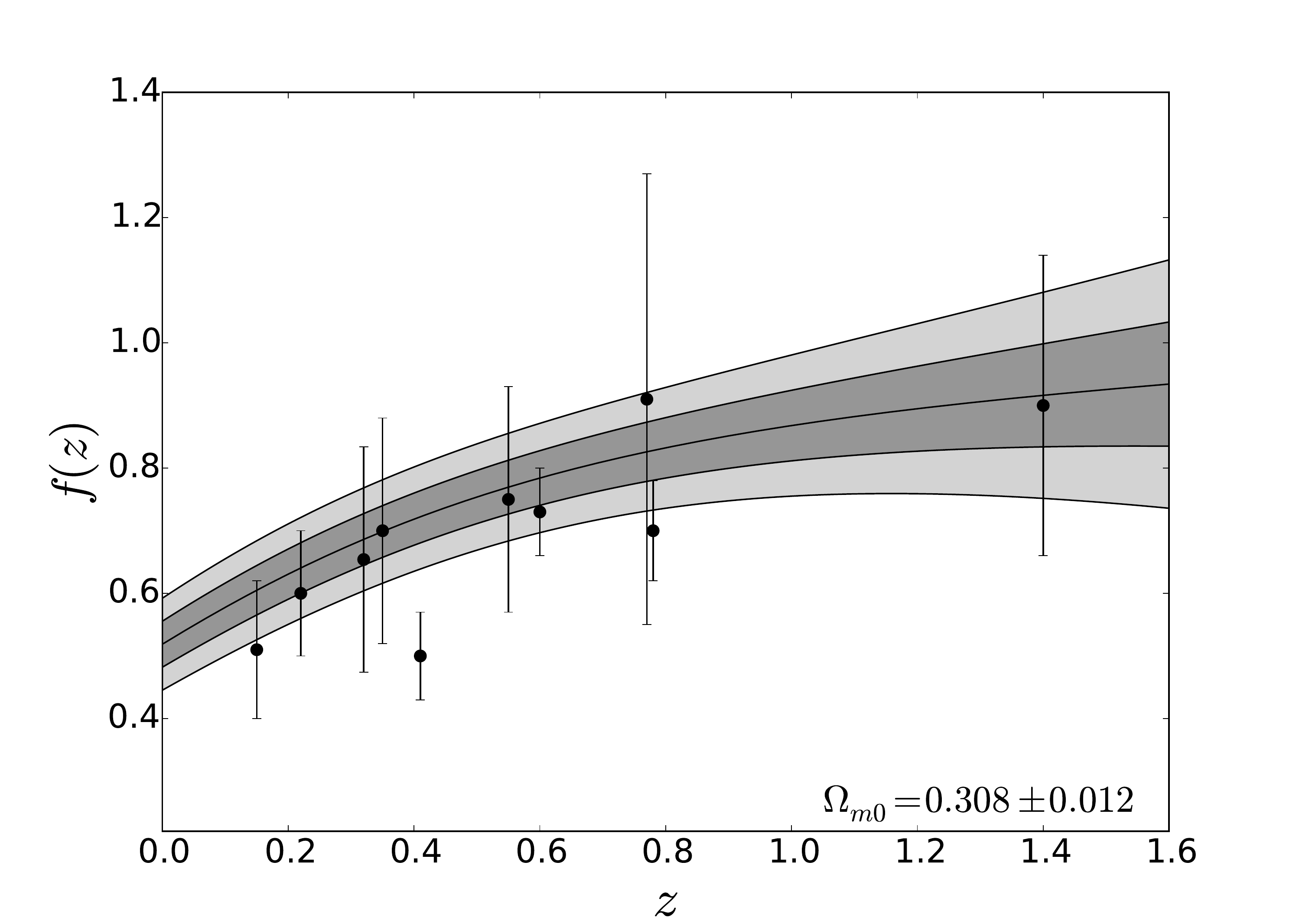}
\hspace{0.1cm}
\includegraphics[width = 8.5cm, height = 6.5cm]{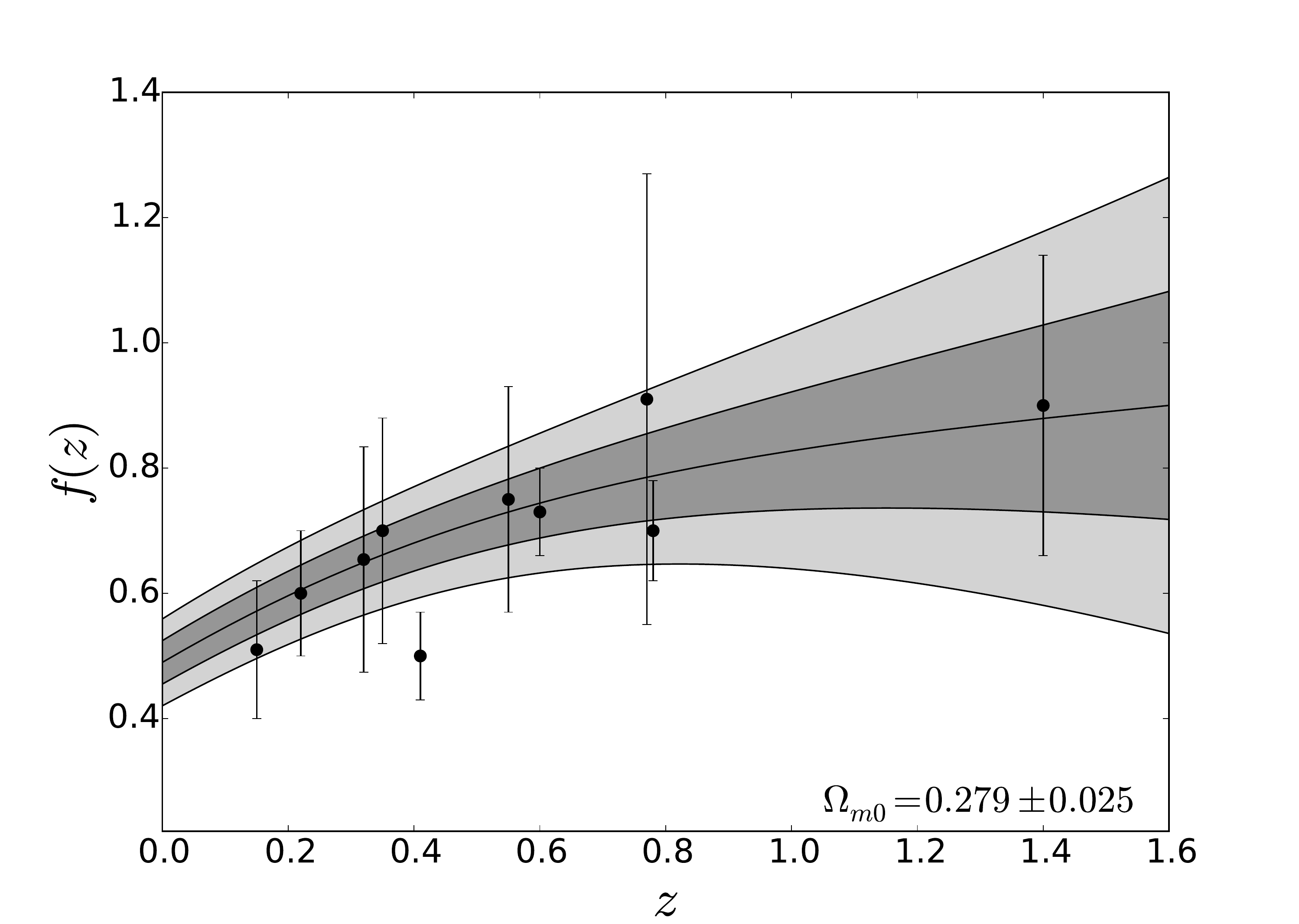}
\caption{Reconstruction of the  growth rate of the matter perturbations. (a) The growth rate  obtained assuming the Planck  $\Omega_{m0}$ value. (b) The same as in the previous panel assuming the WMAP-9  $\Omega_{m0}$ value.   The  data points were taken from Table II of Ref. \cite{GuptaTabela} and the shaded regions  represent the 1$\sigma$ and 2$\sigma$ confidence levels.}.
\label{fz}
\end{figure*}

\begin{figure*}[t]
\includegraphics[width = 8.5cm, height = 6.5cm]{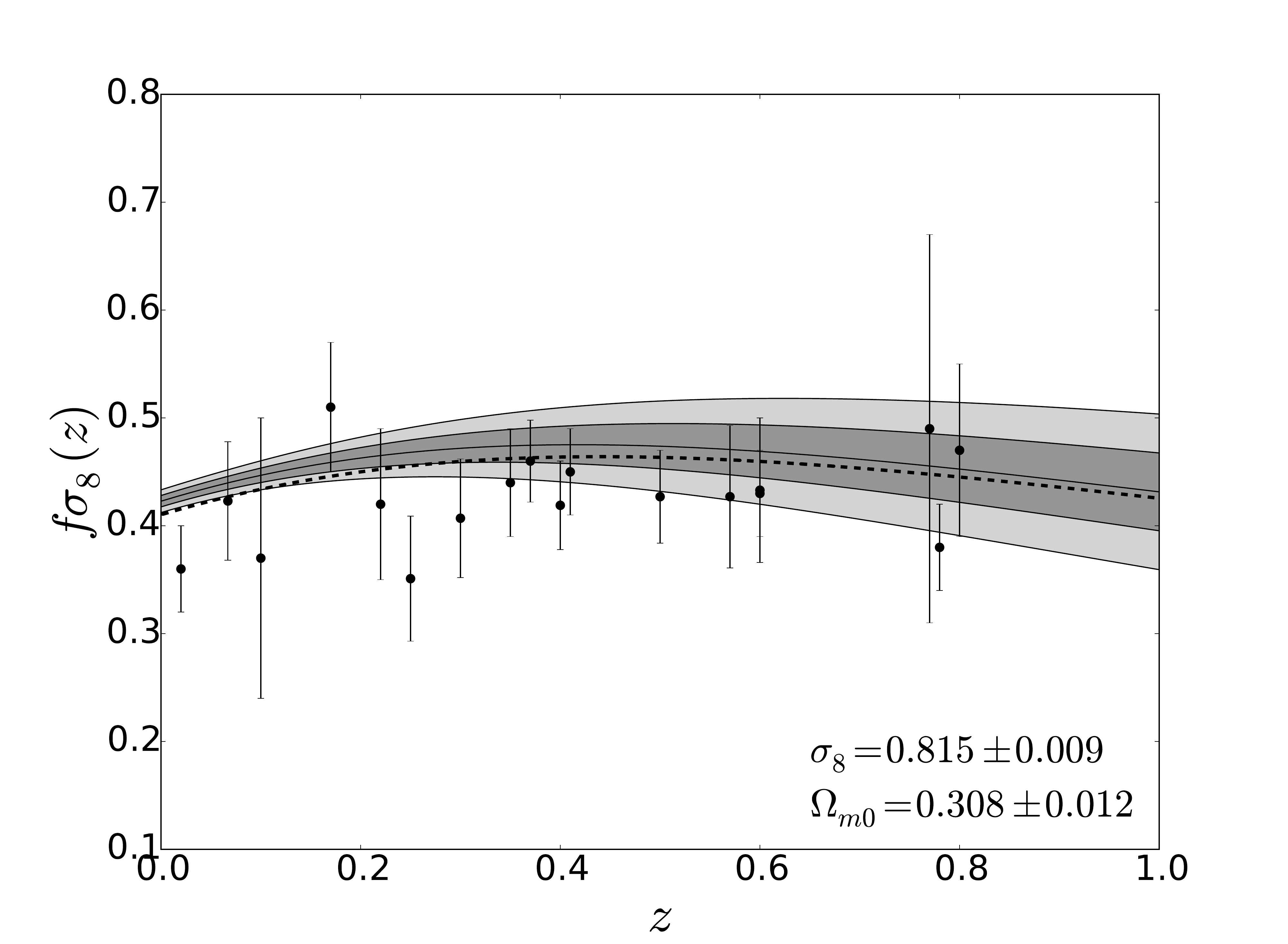}
\hspace{0.1cm}
\includegraphics[width = 8.5cm, height = 6.5cm]{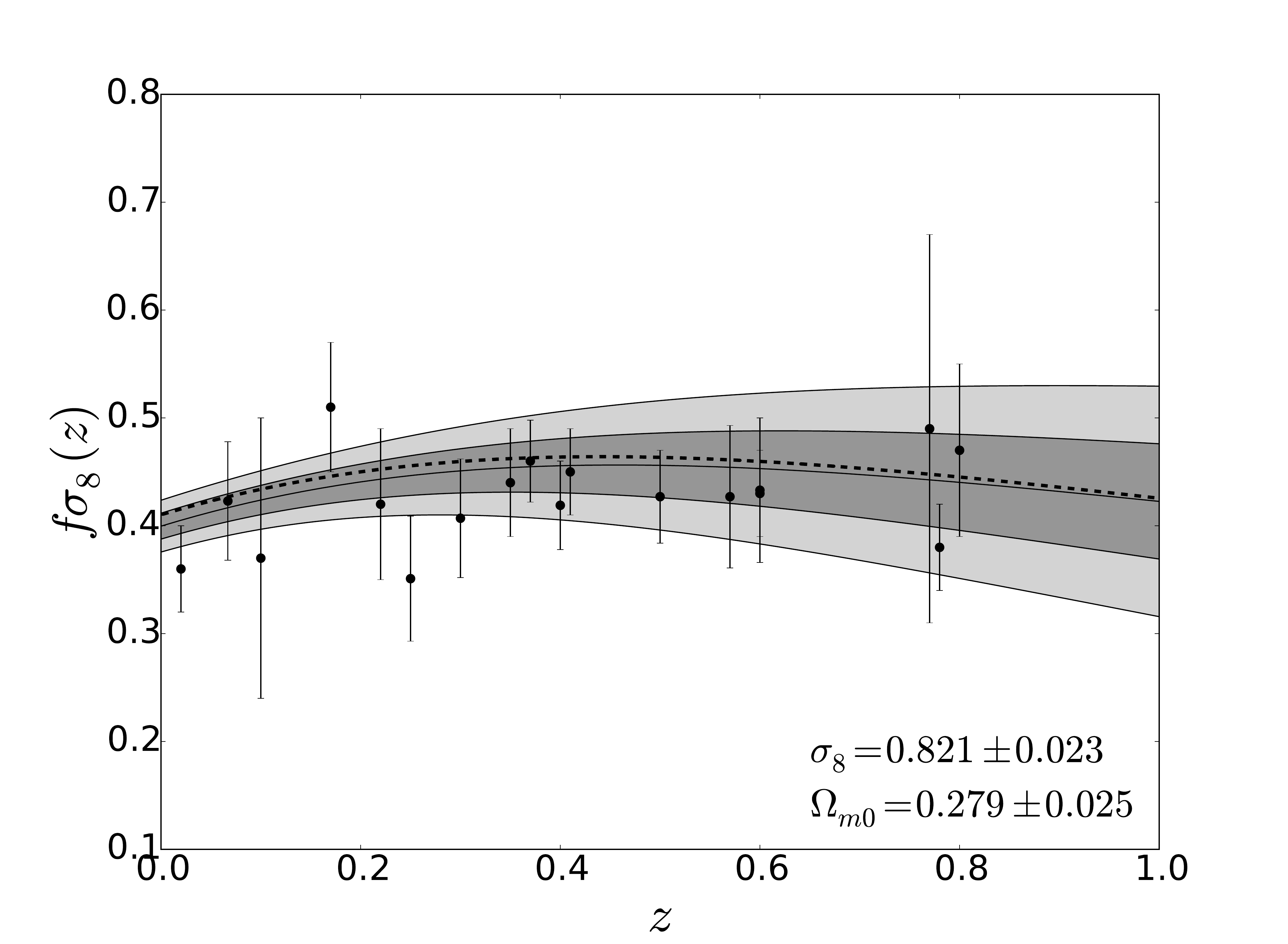}
\caption{The  rate of change of the clustering amplitude, $f\sigma_8(z)$. The solid line corresponds to the reconstruction while the shaded regions  represent the 1$\sigma$ and 2$\sigma$ confidence levels. For comparison, we also show the $\Lambda$CDM prediction (dashed line). (a) The  rate of change of the clustering amplitude  obtained assuming the Planck  $\Omega_{m0}$ and $\sigma_8$ values. (b) The same as in the previous panel assuming the WMAP-9  $\Omega_{m0}$ and $\sigma_8$ values.  The  data points were taken from Table I of Ref. \cite{Nesseristest}.}
\label{fs8z}
\end{figure*}

 \begin{figure*}[t]
\includegraphics[width = 8.5cm, height = 6.5cm]{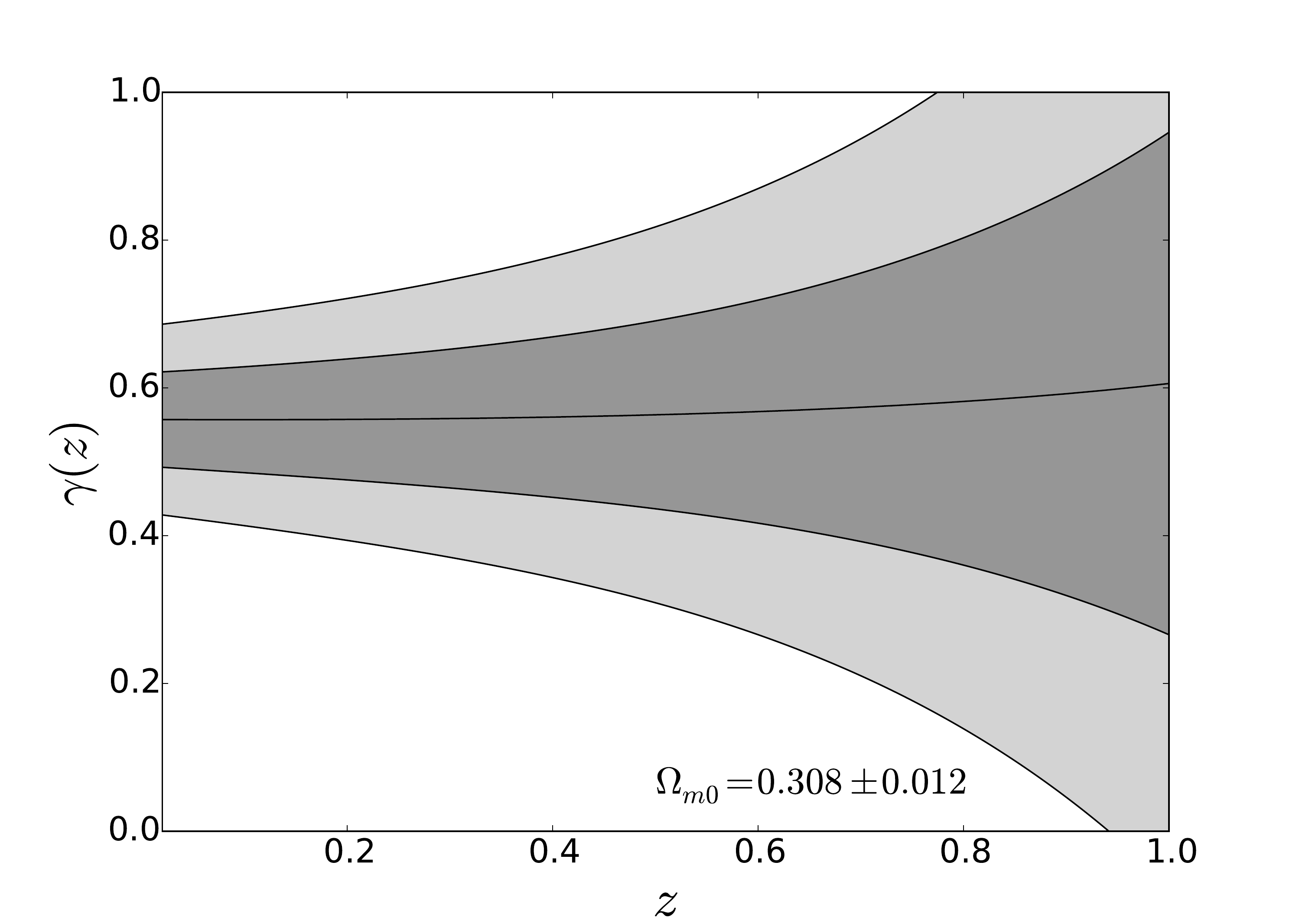}
\hspace{0.1cm}
\includegraphics[width = 8.5cm, height = 6.5cm]{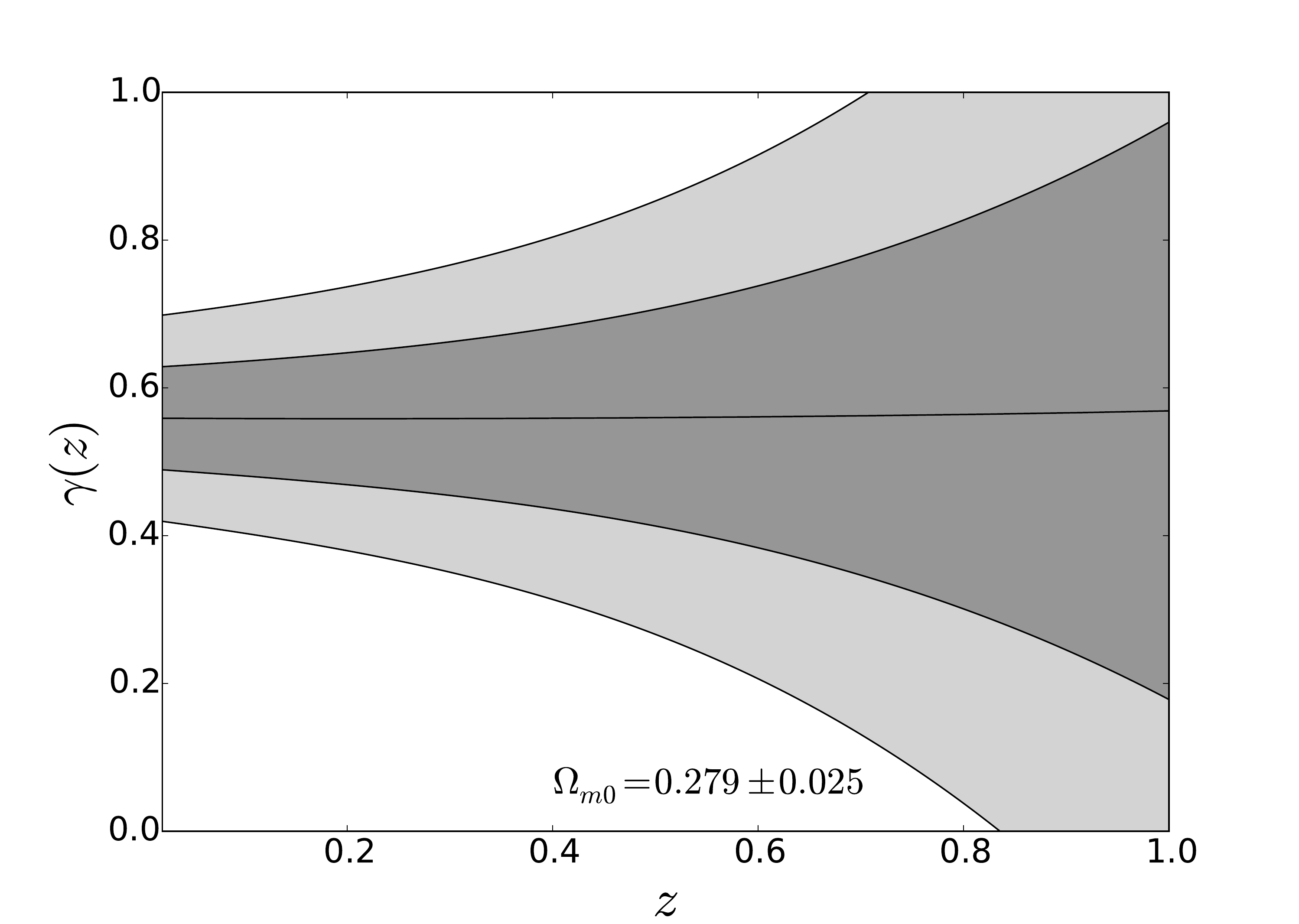}
\caption{The growth index $\gamma(z)$ of matter perturbations. The solid line 
corresponds to the reconstruction from the NPS method whereas the shaded regions  represent the 1$\sigma$ and 2$\sigma$ 
confidence levels. (a) The growth index obtained assuming the Planck  $\Omega_{m0}$ value. (b) The same as in the previous panel assuming the WMAP-9  $\Omega_{m0}$ value.}
\label{gammaz}
\end{figure*}

 \begin{figure*}[t]
\includegraphics[width = 8.5cm, height = 6.5cm]{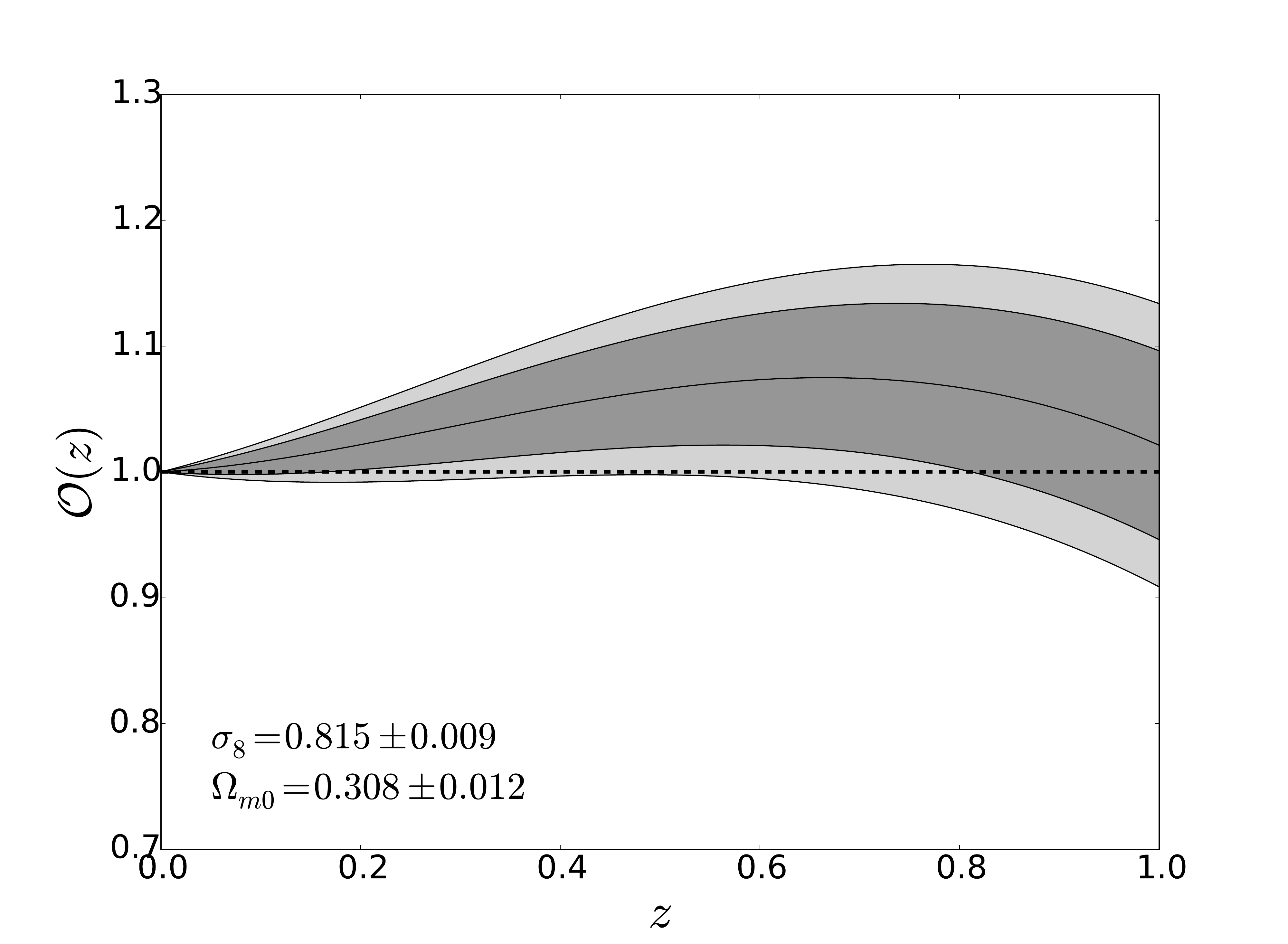}
\hspace{0.1cm}
\includegraphics[width = 8.5cm, height = 6.5cm]{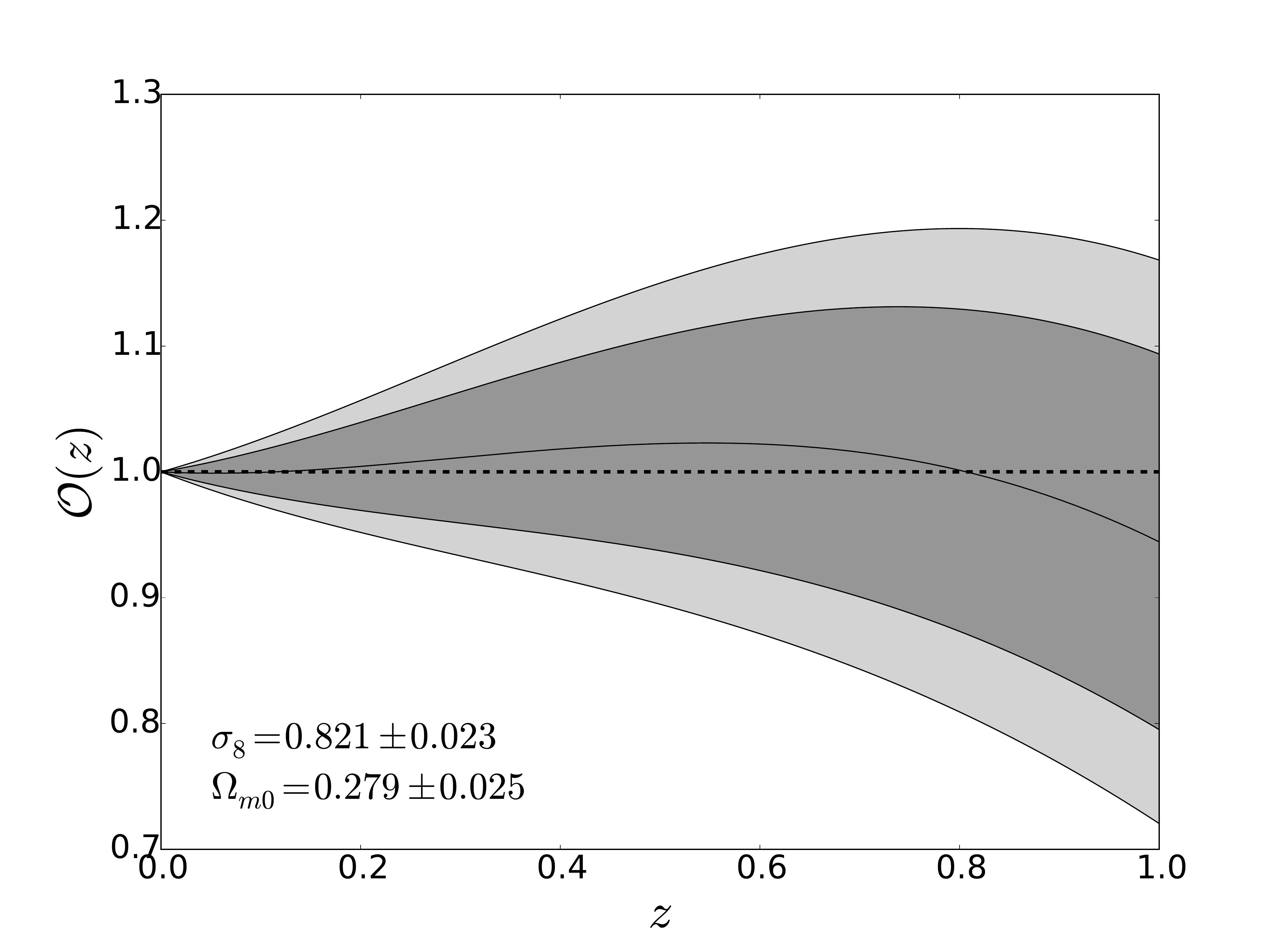}
\caption{Results of the null test ${\cal{O}}(z)$. (a) The evolution of ${\cal{O}}(z)$ obtained assuming the Planck  $\Omega_{m0}$ and $\sigma_8$ values. (b) The same as in the previous panel assuming the WMAP-9  $\Omega_{m0}$ and $\sigma_8$ values. The shaded regions  represent 2$\sigma$ and  3$\sigma$ confidence levels. The dashed line corresponds to the ${\cal{O}}(z)$ value when GR is valid.}
\label{Otestfig}
\end{figure*}
 
\section{Results}
\label{R}

In order to perform our non-parametric reconstruction of the Hubble parameter, 
we apply the  method described in section \ref{NPS} to the cosmic chronometer and high-$z$ quasar data presented in Table \ref{Hdados}. 
The final reconstruction is shown in Fig. \ref{Hzfig}(a) along with the time variation of the scale factor, i.e., $\dot{a}=H(z)/(1+z)$ (Fig. 1(b)). 
It is important to explore this latter quantity because in the following analysis of matter perturbations we assume the existence of a matter-dominated 
epoch and, therefore, the existence of a decelerated phase in the cosmic expansion. In  Fig. \ref{Hzfig}(b), 
the transition redshift corresponds to the minimum of the function
at $z_t\simeq 0.5$ which is in agreement with recent  estimates of this quantity~\cite{jvital,thoven,Moresco2016}. 

In Figs. \ref{Omfig}(a) and \ref{Omfig}(b) we show the 
matter density parameter (Eq.(\ref{om})) obtained from the smoothed $H(z)$ function and two priors for $\Omega_{m0}$ given by 
Planck collaboration \cite{Planck}, $\Omega_{m0}=0.308 \pm 0.012$ , and by WMAP-9 collaboration, $\Omega_{m0}=0.279 \pm 0.025$ \cite{WMAP9}. Using the $H(z)$ and the $\Omega_{m0}$ priors, as mentioned in Sec. \ref{MPE}, 
we also calculate the matter density contrast by solving the integral (\ref{22odes}). 
As discussed above, the evolution of $\delta(z)$ is totally determined by the cosmic expansion and by the matter density content at the present 
day when we fix the integration constant $\delta'_0$. 
From the perturbation theory we know exactly how  is the behaviour of the density contrast in the matter dominated epoch. 
We know that in GR the matter density growth is proportional to $a=1/(1+z)$ when $\Omega_m(z)\simeq 1$, 
which is expected at high-$z$. 
This behaviour is more easily identifiable if we analyze the growth factor at high redshift.
In this regime the behaviour of $g(z)$ must be constant. 
We explore different values for the constant $\delta'_0$ and 
we choose the one when we reach the expected behaviour of the growth factor close to the highest redshift of our sample,
$z=2.34$. The final growth factor function is very sensitive to the $\delta'_0$ value,
therefore we can obtain an accurate estimate of this integration constant. 
The resulting $\delta(z)$ depends on the $\Omega_{m0}$ and hence the correct $\delta'_0$ value also depends on the matter content. 
For the matter density contrast calculated with Planck and WMAP-9 $\Omega_{m0}$ values we infer $\delta'_0=0.519\pm 0.003$ and $\delta'_0=0.49\pm 0.003$, respectively. 

In Fig. \ref{fz} we show the resulting reconstruction of the growth rate. 
In this plot we also display the $f(z)$ measurements  compiled in Ref. \cite{GuptaTabela}.
The previous estimate of the $\delta'_0$ constant is very important in the determination of the $f(z)$. 
From the  alternative expression given by Eq.(\ref{fdef}), in terms of $\delta$ and $H(z)$, we can show that the growth rate at the present epoch and $\delta'_0$ are related trough  $f(0)=-\delta'_0$. 
The other observable, $f\sigma_8(z)$, is shown in Fig. \ref{fs8z}. 
In this case we also need  information about the $\sigma_8$ parameter to determine the function (Eq.(\ref{fs8def})) 
from the reconstructed cosmic expansion. We use the Planck and WMAP-9 collaboration values,  $\sigma_8=0.815\pm0.009$ and $\sigma_8=0.821\pm0.023$, respectively. In Fig. \ref{fs8z}
we also show the $f\sigma_8(z)$ measurements compiled in Ref. \cite{Nesseristest}. 
These data, $f(z)$ and $f\sigma_8(z)$, are usually obtained from data of large-scale galaxy distribution and are derived assuming a fiducial cosmology. 
The compatibility  between the reconstructed quantities from the $H(z)$ function and the 
observational data shows a good agreement  with the standard 
cosmological model, that is, with the hypothesis of a homogeneous and isotropic universe filled with matter and DE fluid covariantly conserved in the framework of  GR (see Sec \ref{MPE}). 
An agreement between this reconstructed approach and the data  was also pointed out in Ref. \cite{Gonzalez}
using Gaussian Processes (GP) to obtain the $H(z)$ function. 
Therefore, we notice that this agreement does not depend on the non-parametric method used to reconstruct the cosmic expansion. 

The last perturbative quantity defined in Sec \ref{MPE}, $\gamma(z)$, is calculated with the  reconstructed growth rate (Fig \ref{fz}) and with the matter density parameter (Fig. \ref{Omfig}) for  both values of $\Omega_{m0}$ from CMB experiments. At $z = 0$, we found $\gamma_0 = 0.57\pm0.13$ $(2\sigma)$ and $\gamma_0 = 0.56\pm0.14$ $(2\sigma)$ for the Planck and WMAP-9 values of $\Omega_{m0}$, respectively. These results are in  good agreement with the results presented in Ref \cite{Gonzalez}.  The reconstruction of the growth index is shown in Fig. \ref{gammaz}. 

Finally, we also apply the NPS method to the  $f\sigma_8(z)$ data displayed in Fig. \ref{fs8z} to obtain the evolution of this quantity in a non-parametric form. With the $H(z)$ and  $f\sigma_8(z)$ observables reconstructed and he $\sigma_8$ and $\Omega_{m0}$ estimates from CMB experiments, we calculate the perturbative null test ${\cal{O}}(z)$, discussed in Sec. (\ref{nt}). The results are shown  in Fig. \ref{Otestfig}. We note that the non-parametric reconstruction of the quantity ${\cal{O}}(z)$ shows  a tension between the standard model prediction and the data at $\simeq 2\sigma$ level for the results obtained with Planck collaboration priors. On the other hand, the same reconstruction performed with WMAP-9 priors is consistent with the standard cosmology.     

\section{Conclusions}
\label{C}
In this work we have performed a reconstruction of perturbative quantities using a non-parametric smoothing method applied to current data of the cosmic expansion. The fact that both the method of reconstruction and the data set are model-independent reduces the possibility of biased results. 

Analyzing the growth rate, $f(z)$, and the  rate of change of the clustering amplitude, $f\sigma_8(z)$, calculated by solving the perturbation Eqs. (\ref{22odes}), (\ref{fdef}) and (\ref{fs8def}) and comparing with their observational estimates, we have not found any deviation from the predictions of the standard cosmology. This is consistent with the results reported  in Ref. \cite{Gonzalez}, where a similar analysis was performed using a different reconstruction method. 

The previous result is confirmed by calculating the ${\cal{O}}(z)$ test with the reconstructed $H(z)$ and $f\sigma_8(z)$ evolutions via NPS using the values of $\sigma_8$ and $\Omega_{m0}$  from the WMAP-9 collaboration. However, the result is not the same when we calculate the perturbative null test using the corresponding values from the Planck collaboration. In this case, we have found a violation of the null test at $\sim 2\sigma$ level, which is not evident in Figs. \ref{fz}(a) and \ref{fs8z}(a). It shows the high sensitivity of the ${\cal{O}}(z)$ test compared with the perturbative quantities.  We have also performed the null test using different smoothing scales to reconstruct the $f\sigma_8(z)$. However, in these cases, there is a tension for the both sets of $\Omega_{m0}$ and $\sigma_8$ CMB priors. Such result is similar to the one discussed in Ref. \cite{Nesseristest}, where a violation or not of the test depends on the binning of the data.   

Finally, we have calculated the growth index and found that it is more effectively constrained at $z=0$, $\gamma_0 = 0.57\pm0.13
$ $(2\sigma)$. Such result is compatible with the $\Lambda$CDM expected value ($\gamma=6/11$) and its first derivative $\gamma'_0\simeq-0.015$ \cite{GannoujiPolarski}. We also emphasize that the growth index estimate obtained in this work using NPS method is in good agreement with the results presented in Ref. \cite{Gonzalez} using GP method.

\begin{acknowledgments}
JEG, JSA and JCC thank CNPq, CAPES and FAPERJ (Brazilian agencies) for the grants under which this work was carried out. 
\end{acknowledgments}

\end{document}